\documentclass[prl,twocolumn,aps,letterpaper, preprintnumbers,superscriptaddress]{revtex4-2}

\usepackage{graphicx}% Include figure files
\usepackage{dcolumn}% Align table columns on decimal point
\usepackage{bm}% bold math

\usepackage{amsthm,amsmath,amssymb}
\usepackage{mathrsfs}
\usepackage{appendix}
\usepackage{amstext} 
\usepackage{url}
\usepackage{float}
\usepackage[usenames,dvipsnames]{color}
\usepackage[colorlinks=true,citecolor=blue,urlcolor=black]{hyperref}% add hypertext capabilities

\usepackage{booktabs}
\usepackage[linesnumbered,algoruled,boxed,lined]{algorithm2e} 
\usepackage{siunitx}
\usepackage{makecell}

\def\ket#1{\mathinner{|{#1}\rangle}}
\def\ketbra#1#2{\mathinner{|{#1}\rangle\langle{#2}|}}

\theoremstyle{plain}

\begin{document}  

\title{Source-independent quantum key distribution without pre-sending entanglement}

\author{Rong-Zheng Liu}
\author{Hua-Lei Yin}\email{hlyin@ruc.edu.cn}
\affiliation{School of Physics and Key Laboratory of Quantum State Construction and Manipulation (Ministry of Education), Renmin University of China, Beijing 100872, China}

\date{\today}

\begin{abstract}
Quantum key distribution (QKD) theoretically offers information-theoretic security. The prevailing approach is the prepare-and-measure BB84 protocol, which implements QKD using conventional laser rather than single-photon source via the decoy-state method. However, side-channel attacks targeting sources severely threaten system security. Despite extensive efforts, including fully passive scheme, this vulnerability persists even with perfect single-photon source. Here, we propose a source-independent (SI) QKD protocol that resolves all known and unknown source-side attacks without pre-sending entanglement source. Aligning with advances in quantum light sources, our protocol simultaneously doubles the transmission distance while remaining robustness against imperfection of source. Theoretical analysis shows that non-classical light source provides practical security advantages unattainable with conventional laser. 
\end{abstract} 

\maketitle
 
Quantum key distribution (QKD) promises information-theoretic security via quantum laws~\cite{lo2014secure,yin2016measurement}. However, discrepancies between theory and practice lead to security loopholes and eavesdropping risks~\cite{zhao2008quantum,lydersen2010hacking,jain2011device,xu2020secure}. To bridge the gap, device-independent QKD protocols have been proposed, but the demanding requirements for transmission efficiency render them impractical for long-distance application~\cite{zhang2022device,nadlinger2022experimental}. Subsequently, the development of measurement-device-independent QKD~\cite{lo2012measurement} and its variants based on asynchronous two-photon interference~\cite{xie2022breaking,zeng2022mode} and single-photon interference~\cite{Lucamarini2018overcome} have eliminated the security assumption in measurement devices, thereby resolving all detector-side attacks~\cite{braunstein2012side}. 
\begin{table}[b]
    \renewcommand{\arraystretch}{1.2}
    \caption{Source-side attacks. }
    \centering
    \begin{tabular*}{\linewidth}{l l}
    \hline
    \hline
    ~~~~~~~~\textbf{Attack Type} & ~~~~~~~~~~\textbf{Reference} \\ 
    \hline
    ~~~~~~~~Photon-number splitting     & ~~~~~~~~~~Ref.~\cite{brassard2000limitations,Lutkenhaus2002quantum} \\
    ~~~~~~~~Light-injection             & ~~~~~~~~~~Ref.~\cite{gisin2006trojan,lucamarini2015practical,guo2025discrete,huang2020laser,pang2020hacking,ponosova2022protect,peng2024security} \\
    ~~~~~~~~Wavelength-dependent        & ~~~~~~~~~~Ref.~\cite{li2011attacking,peng2025practical} \\
    ~~~~~~~~Phase-remapping             & ~~~~~~~~~~Ref.~\cite{fung2007phase,xu2010experimental} \\
    ~~~~~~~~Hidden-multidimensional     & ~~~~~~~~~~Ref.~\cite{tang2013source,gnanapandithan2025hidden} \\
    ~~~~~~~~State-flaw                  & ~~~~~~~~~~Ref.~\cite{tamaki2016decoy,wang2018finite,pereira2020quantum}\\
    \hline
    \hline
    \end{tabular*}
    \label{tab:attcks}
\end{table}
Therefore, vulnerabilities originating from the source have emerged as a critical security concern~\cite{brassard2000limitations,Lutkenhaus2002quantum,gisin2006trojan,lucamarini2015practical,guo2025discrete,huang2020laser,pang2020hacking,ponosova2022protect,peng2024security,li2011attacking,peng2025practical,fung2007phase,xu2010experimental,tang2013source,gnanapandithan2025hidden,tamaki2016decoy,wang2018finite,pereira2020quantum}, as summarized in Table~\ref{tab:attcks}. 

Despite significant progress made in solving source-side vulnerabilities including loss-tolerant protocol~\cite{tamaki2014loss,tang2016experimental,pereira2019quantum,pereira2020quantum,gu2022experimental,li2025quantum}, side-channel-secure scheme~\cite{wang2019practical,zhang2022experimental}, and fully passive scheme~\cite{wang2023fully,lu2023experimental,hu2023proof}, these methods only partially characterize the imperfection of source flaws. Unknown attacks could still pose substantial threats to the security of QKD systems. For example, recently identified hidden multidimensional modulation side channels introduce unnoticed security loopholes and severely limit the maximum secure transmission distance~\cite{gnanapandithan2025hidden}. Therefore, the development of protocols independent of the source remains an urgent and fundamental challenge. 

\begin{figure}[t]
\centering
\includegraphics[width=\linewidth]{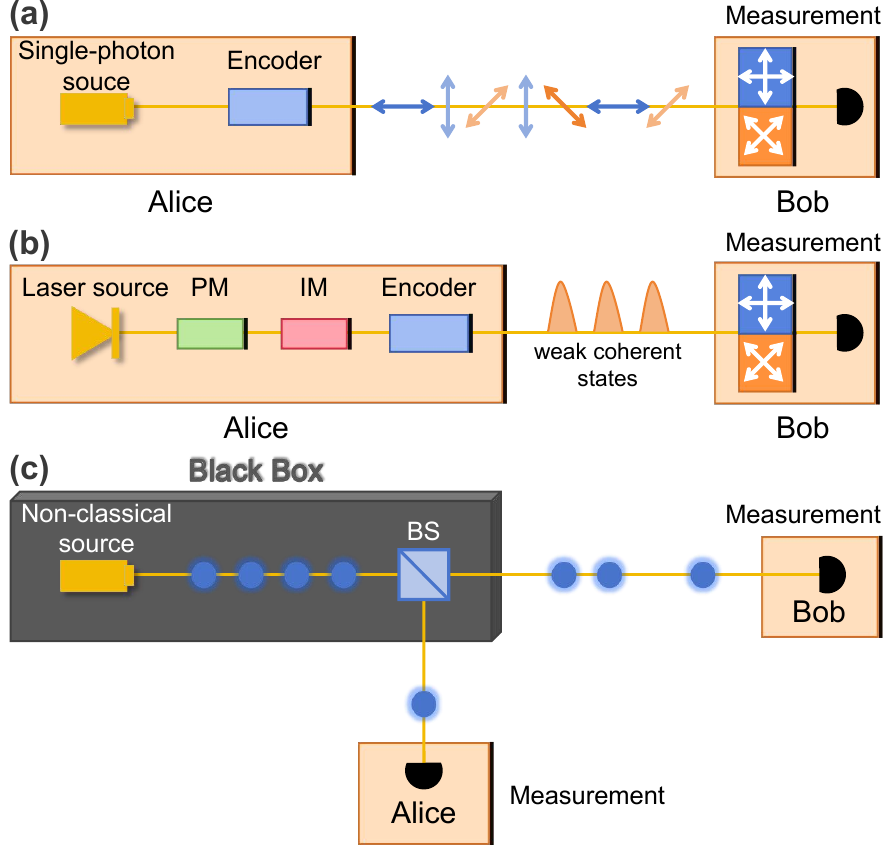}
    \caption{\textbf{Protocols comparison.}
    (a) Single-photon based and (b) WCS-based BB84 protocols: both the source and measurement devices must be trusted.
    (c) SI-QKD protocol: The non-classical source generation and distribution are modeled as an untrusted black box, while the local measurements at Alice and Bob are trusted. PM: phase modulator, IM: intensity modulator, BS: beam splitter. }
    \label{fig:protocol comparison}
\end{figure}

\begin{figure*}[t]
    \centering
    \includegraphics[width=0.9\linewidth]{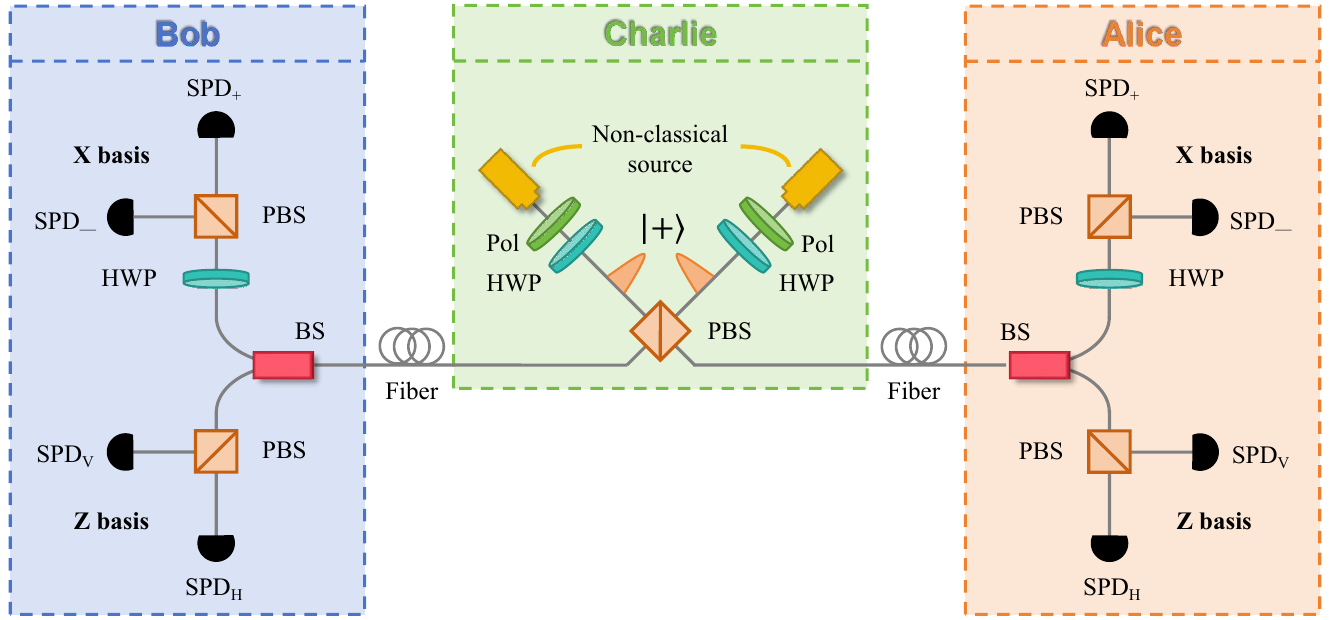}
    \caption{\textbf{The polarization scheme of SI-QKD protocol.}
    Charlie uses two independent non-classical sources that each emits a sequence of optical pulses. Each photon is prepared in the $\ket{+}$ state by a polarizer (Pol) and half-wave plate (HWP). Pulses from two sources interfere at a polarizing beam splitter (PBS) and distributed to receivers. Alice and Bob use a local beam splitter (BS) to randomly select the measurement basis, followed by polarization analysis in either the rectilinear (\textit{Z}) basis or the diagonal (\textit{X}) basis. Measurement outcomes are recorded by four single-photon detectors (SPDs) corresponding to polarization bases.}
    \label{fig:SI-QKD scheme}
\end{figure*}

Recently, non-classical quantum sources have attracted growing interest as an important resource for quantum networks~\cite{couteau2023applications,couteau2023application}. Non-classical light sources here refer to quantum sources that emit high-purity single-photon states, such as single-photon source (SPS)~\cite{senellart2017high,garcia2021semiconductor} and Schr{\"o}dinger cat state~\cite{wineland2013nobel,li2017cat,he2023fast}. As shown in Fig.~\ref{fig:protocol comparison}(a), the integration of SPS into prepare-and-measure BB84~\cite{bennett1984quantum} protocols has achieved remarkable advancement, for example, high-rate intercity QKD experiment~\cite{yang2024Highratea}, surpassing weak coherent state (WCS) rate limit~\cite{zhang2025experimental}, implementation with truncated decoy-state or heralded-purification protocol~\cite{bloom2025decoy} and long-distance QKD with SPS~\cite{morrison2023singleemitter}. However, these approaches are still vulnerable to source-side channel threats and have difficulty outperforming WCS-based decoy-state BB84~\cite{hwang2003quantum,wang2005beating,lo2005decoy}, as shown in Fig.~\ref{fig:protocol comparison}(b). 

In this Letter, we propose a source-independent (SI) QKD protocol based on non-classical source, which is immune to all source-side attacks and shown in Fig.~\ref{fig:protocol comparison}(c). Our protocol highlights the essential distinction between laser and non-classical source in realizing SI security. Simultaneously, our protocol doubles the secure distance of single-photon BB84~\cite{morrison2023singleemitter} and surpasses the WCS-based decoy-state BB84~\cite{yin2020tight}. These results establish the non-classical source as a practical resource for source-independent quantum cryptography.

As illustrated in Fig.~\ref{fig:SI-QKD scheme}, our SI-QKD protocol adopts a polarization scheme, where quantum information is carried by the polarization degree of freedom of single photons. In addition, our protocol can also be realized in alternative implementations, such as the time-bin scheme and the phase scheme. Charlie utilizes two non-classical sources with high single-photon purity ($g^{(2)}(0)\ll 1$), such that each pulse approximates the single-photon state $|1\rangle$. Alice and Bob then perform measurements either in the rectilinear (\textit{Z}) or diagonal (\textit{X}) basis. Notably, quantum correlations between communication parties are based on the indivisibility of single-photon and post-selected measurements, without requiring pre-prepared entangled photon pairs. The protocol proceeds in the following five steps.

\textbf{Step 1.}—An untrusted third party, Charlie, employs two independent non-classical sources with $g^{(2)}(0)\ll 1$, each emitting a sequence of optical pulses in time slots $i=1,2,...,N$. Through a polarizer and a half-wave plate, each pulse is initialized in the diagonal polarization state $\ket{+}=(\ket{H}+\ket{V})/\sqrt{2}$. The sequences of states from two sources are synchronized and injected into two input ports, $c$ and $d$, of the polarizing beam splitter (PBS). The interference progress can be expressed as
\begin{equation}
    \ket{+}_{c}\ket{+}_{d}\xrightarrow{\rm PBS}\frac{1}{2}\left(\ket{HH}_{ab}+\ket{VV}_{ab}+\ket{HV}_{aa}+\ket{VH}_{bb}\right),
\end{equation}
where modes $a$ and $b$ denote the output ports directed toward Alice and Bob, respectively. The resulting states are distributed to communication parties through insecure quantum channels.

Notably, the aforementioned architecture can be realized equivalently using a single non-classical source with an active optical switch. The source generates a sequence of independent pulses where each photon is prepared in the $\ket{+}$ state. The periodic optical switch then routes the pulses according to the parity of their temporal indices. By matching the difference in path length to the pulse repetition interval, adjacent pulses are synchronously injected into the input ports of the PBS and interfere with each other. While an active optical switch is used here for photon routing, it can be replaced by a passive BS to reduce experimental complexity, with the cost of a 50\% reduction in effective counts.

\textbf{Step 2.}—Upon receiving the quantum signals, Alice and Bob independently perform polarization measurements in either the rectilinear basis $\{\ket{H},\ket{V}\}$ or the diagonal basis $\{\ket{+},\ket{-}\}$. With a local beam splitter to passively but randomly select the measurement basis, Alice and Bob record all detection events and corresponding basis. In the rare case that both detectors within the same basis click simultaneously, one of them is randomly chosen as valid. 

\textbf{Step 3.}—For all detections, Alice and Bob publicly announce their measurement bases and retain only those events detected in the same basis. In the rectilinear basis, measurement outcomes $\ket{H}$ and $\ket{V}$ are mapped to bit values 0 and 1 of the \textit{Z} basis, respectively. In the diagonal basis, the states $\ket{+}$ and $\ket{-}$ are assigned bit values 0 and 1 of the \textit{X} basis. 

\textbf{Step 4.}—Alice and Bob utilize the bit string obtained from the sifted \textit{Z}-basis outcomes as the raw key. They disclose all data of the \textit{X} basis to obtain the bit error rate $E_{x}$, and then estimate the upper bound of the phase error rate $\phi_{z}$ in the \textit{Z} basis. 

\textbf{Step 5.}—By employing error correction and privacy amplification, the finite secret key length can be written as
\begin{equation}
    \ell = n_{z}[1-h(\phi_{z})]-\lambda_{\rm{EC}}-\log_2\frac{2}{\varepsilon_{\rm{cor}}}-2\log_2\frac{1}{\varepsilon_{\rm{sec}}},
    \label{eq-l}
\end{equation}
where $n_{z}=Np_{z}^2Q_{z}$ is the number of \textit{Z}-basis detection events, with the measurement probability $p_{z}$ and the total gain $Q_{z}$. $\lambda_{\rm{EC}}=n_{z}fh(E_{z})$ is the information leaked during error correction, where $f$ is the error correction efficiency, $E_{z}$ is the qubit error rate of the \textit{Z} basis, and $h(x)=-x\log_2x-(1-x)\log_2(1-x)$ is the binary Shannon entropy function. $\varepsilon_{\rm{sec}}$ and $\varepsilon_{\rm{cor}}$ are the security coefficients regarding the secrecy and correctness, respectively. Then the finite secure key rate (SKR) can be defined as $r=\ell/N$. The details of the calculation and the security analysis are given in supplemental document.

With the indivisibility of single photon, local measurements and post-selection method can yield correlations between distant parties without the need of pre-distributed entanglement. In contrast, the WCS emitted from a classical laser source can be written as $\ket{\alpha}^{+}=\ket{\frac{\alpha}{\sqrt{2}}}^{H}\otimes\ket{\frac{\alpha}{\sqrt{2}}}^{V}$. The interference at the PBS can be described by
\begin{equation}
\ket{\alpha}_{c}^{+}\otimes\ket{\alpha}_{d}^{+}\xrightarrow{\rm PBS}\ket{\alpha}_{a}^{+}\otimes\ket{\alpha}_{b}^{+},
\end{equation}
where $\ket{\alpha}_{c}^{+}$ denotes the $\ket{+}$-polarized coherent state in port $c$. The output state remains a separable product of independent states, which can neither generate quantum correlation nor support the source-independent security framework.

\begin{figure}[t]
\centering
\includegraphics[width=\linewidth]{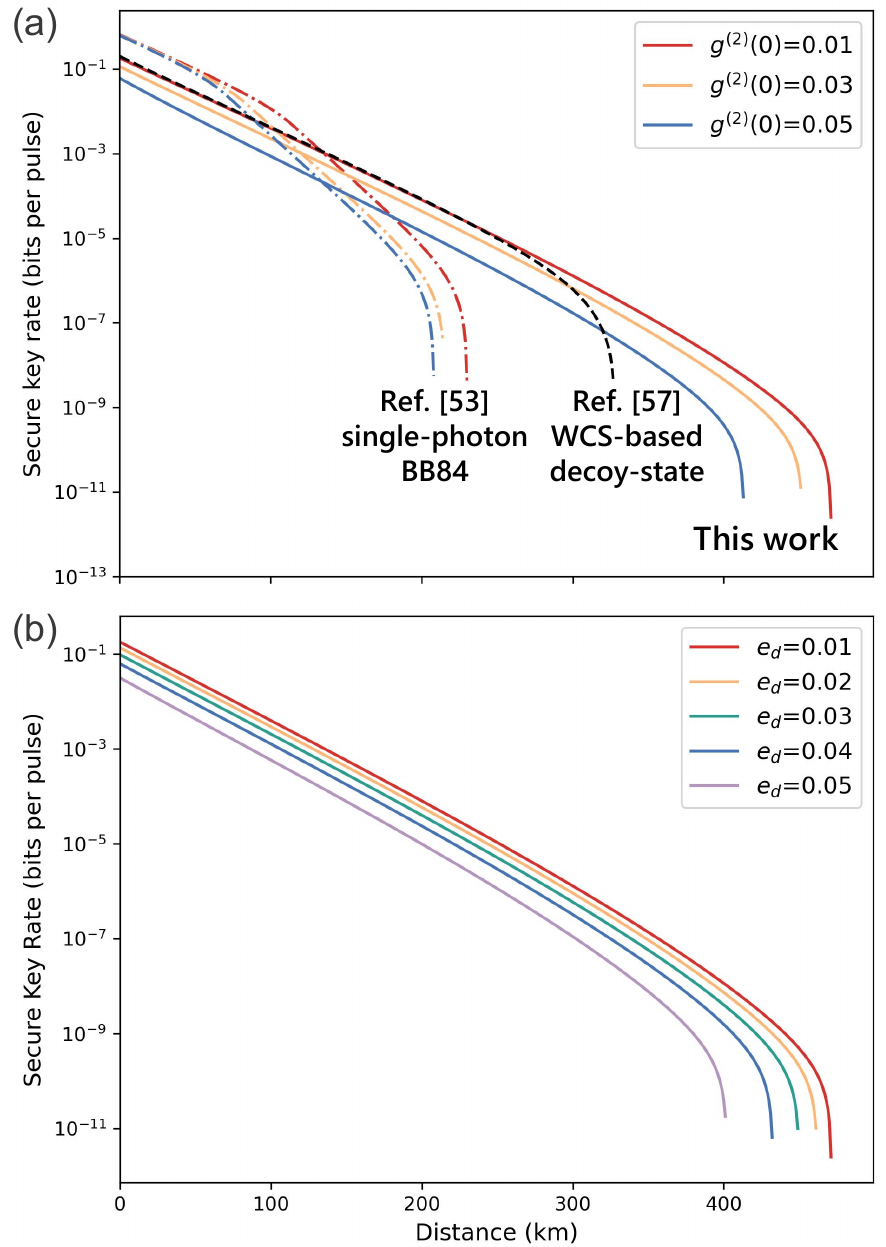}
    \caption{\textbf{Comparison of finite secure key rate.}
    (a) Finite SKR comparison between our polarization SI-QKD protocol (solid lines), the single-photon BB84 protocol~\cite{morrison2023singleemitter} (dash-dotted lines), and the WCS-based decoy-state BB84 protocol~\cite{yin2020tight} (dashed lines). The parameters are set as $e_d=0.01$, $p_d=10^{-7}$ and $N=10^{12}$.   
    (b) Finite SKR for various misalignment error $e_d$ with fixed parameters $g^{(2)}(0)=0.01$. The transmission distance denotes the fiber length between Alice and Bob.}
    \label{fig:SKR-pol}
\end{figure}

The proposed protocol is applicable to non-classical sources with high single-photon purity. For performance benchmarking, we simulate the protocol with pure states from SPS. The photon number distribution, $\{p_n\}=\{p_0,p_1,p_2\}$, is characterized by the average photon number $\langle n\rangle$ and the second-order correlation function $g^{(2)}(0)$~\cite{morrison2023singleemitter}, which serves as the primary metric for multi-photon contribution. We compare the finite SKR of our SI-QKD protocol against the single-photon BB84 protocol~\cite{morrison2023singleemitter} and the WCS-based decoy-state BB84 protocol~\cite{yin2020tight}, as shown in Fig.~\ref{fig:SKR-pol}. The average photon number $\langle n\rangle$ and the $\mathit{Z}$-basis measurement probability $p_z$ are optimized at every distance to maximize the SKR. In the simulation, we set the data size as $N=10^{12}$, the error correction efficiency as $f=1.16$, and the channel attenuation coefficient as $\alpha=0.16~\rm dB/km$. The detector efficiency and dark count rate are $\eta_{\rm{det}}=0.8$ and $p_d=10^{-7}$, and the misalignment error rate is $e_{d}=0.01$.

To evaluate the performance of our protocol, we compare the finite SKR of our SI-QKD protocol against the single-photon BB84 protocol~\cite{morrison2023singleemitter} and the WCS-based decoy-state BB84 protocol~\cite{yin2020tight}, as shown in Fig.~\ref{fig:SKR-pol}(a). The simulation results reveal that, compared to the single-photon BB84 protocol~\cite{morrison2023singleemitter}, our protocol achieves a higher SKR beyond 140 km and extends the transmission distance to over 400 km. The advantage of the transmission distance in our protocol comes from the suppression of noise in the coincidence measurement process, which improves the signal-to-noise ratio. Meanwhile, it maintains robustness against source imperfections as $g^{(2)}(0)$ increases. In comparison with the WCS-based decoy-state BB84 protocol~\cite{yin2020tight}, our scheme exhibits longer maximum secure distance under realistic parameter setting. Fig.~\ref{fig:SKR-pol}(b) further investigates the system's tolerance to the misalignment error $e_d$. Notably, the protocol maintains a secure transmission distance of 400 km even with $e_d=5\%$, confirming its robust adaptability to environmental noise and channel disturbances.

These results demonstrate that the proposed protocol effectively leverages the advantage of non-classical source over the classical laser source in both security and transmission distance. Our protocol not only eliminates the secure assumption on the source, but also overcomes the difficulty of surpassing the WCS-based decoy-state BB84 protocol within the prepare-and-measure framework. The practical feasibility of the proposed protocol primarily hinges on ensuring photon indistinguishability and maintaining quantum channel stability, which are discussed in supplemental document. 

In summary, we present a SI-QKD protocol which harnesses the quantum properties of non-classical source and eliminates the security assumption at source side. The numerical results demonstrate that, under realistic conditions, our protocol achieves longer transmission distance than both the traditional BB84 protocol using SPS~\cite{morrison2023singleemitter} and the decoy-state BB84 protocol using WCS~\cite{yin2020tight}. Notably, the secure distance can exceed 400 km while maintaining robustness against source imperfection and hardware noise. These findings highlight that non-classical quantum sources constitute a distinct resource enabling source-independent security and improving system performance. 

Further efforts may focus on developing hybrid architectures that combine SI-QKD with other quantum communication models~\cite{hua2025experimental,lu2025repeater,xiao2025experimental} to enhance security and efficiency. We remark that our SI-QKD protocol does not surpass the repeaterless key rate bound. Investigating how to develop SI protocols that overcome this limit remains a crucial research direction. Moreover, our results emphasize the practical relevance of non-classical source as a vital resource for quantum networks, providing a promising and practical pathway toward long-distance, high-security quantum communication.

\section{Acknowledgments}
This work is supported by the National Natural Science Foundation of China (Nos. 12522419, U25D8016 and 12274223).

\appendix
\section{Appendix A: Security analysis}

In this section, we analyze the security of the proposed source-independent quantum key distribution (SI-QKD) protocol. We assume that the quantum source is fully untrusted and may be controlled by an adversary (Eve), who can prepare and distribute arbitrary quantum states to the receivers. The only trusted components are the local measurements of Alice and Bob, which are assumed to be well-characterized and perform independent measurements in two complementary bases.

According to the entropy uncertainty relation~\cite{tomamichel2011uncertainty}, for two complementary measurements $\mathit{Z}$ and $\mathit{X}$, the generalized uncertainty relation can be written as
\begin{equation}
    H_{\min}^{\varepsilon}(Z_{A}|E) + H_{\max}^{\varepsilon}(X_{A}|B) \geq q,
\end{equation}
where $q=\log_2 \frac{1}{c}$ with $c=\max_{z,x}|\langle z|x\rangle|^2$ quantifying the incompatibility between the two measurement bases. For mutually unbiased bases, one has $q=1$. The smooth min-entropy $H_{\min}^{\varepsilon}(A|B)$ denotes the number of bits contained in $A$ that are $\varepsilon$-close to being uniformly distributed and independent of the quantum system $B$, with the smoothing parameter $\varepsilon\ge0$. The smooth max-entropy $H_{\max}^{\varepsilon}(A|B)$ denotes the number of bits needed to reconstruct the value $A$ using the quantum system $B$ with failure probability $\varepsilon$.

In our protocol, the measurement outcomes in the $\mathit{Z}$ basis are used to generate the raw key, while those in the $\mathit{X}$ basis are used for parameter estimation. After performing error correction and privacy amplification, the length of shared key between Alice and Bob can be calculated by
\begin{equation}
    \ell\approx H_{\min}^{\varepsilon}(Z_A|E)-H_{\max}^{\varepsilon}(Z_A|B).
\end{equation}
Specifically, $H_{\min}^{\varepsilon}(Z_A|E)$ represents the min entropy that Eve has about Alice's raw key $Z_A$, and $H_{\max}^{\varepsilon}(Z_A|B)$ represents the max entropy that Bob has about $Z_A$. Employing the uncertainty relation, the information accessible to Eve can be calculated with the max entropy measured in the complementary basis between Alice and Bob: 
\begin{equation}
\begin{aligned}
    H_{\min}^{\varepsilon}(Z_A|E)&\ge qn-H_{\max}^{\varepsilon}(X_A|B)\\
    &=n[1-(e_p)],
\end{aligned}
\end{equation}
where $n$ is the length of raw key and $e_p$ is the phase error rate. The correlation between the two raw keys quantifies the cost of error correction with $H_{\max}^{\varepsilon}(Z_A|B)\le nh(e_b)$, where $e_b$ is the bit error rate. Therefore, the length of generated secure key can be derived by
\begin{equation}
    \ell\ge n[1-h(e_p)-h(e_b)].
\end{equation} 
Accounting for error correction and finite-size effect, the finite secret key length that can be extracted from $n_{z}$ sifted bits is given by~\cite{tomamichel2012Tight, yin2019Finitekey}
\begin{equation}
    \ell = n_{z}[1-h(\phi_{z})-fh(E_z)]-\log_2\frac{2}{\varepsilon_{\rm{cor}}}-2\log_2\frac{1}{\varepsilon_{\rm{sec}}},
\end{equation}
where $\phi_{z}$ and $E_z$ denote the phase error rate and bit error rate of $\mathit{Z}$ basis, respectively, and $f$ is the error correction efficiency.

Notably, the above security proof does not rely on any assumption about the state preparation process. The security is guaranteed by the independent measurements in complementary bases and the entropy uncertainty relation. Therefore, our protocol establishes a source-independent security framework without requiring pre-distributed entanglement. The entropy uncertainty relation ensures that any attempt by Eve to gain information about the $\mathit{Z}$-basis outcomes necessarily induces disturbances in the $\mathit{X}$ basis.

\section{Appendix B: Derivation for the single-photon BB84 protocol}
Here, we provide the derivation of the single-photon BB84 protocol following Ref.~\cite{morrison2023singleemitter}. Assuming that the multi-photon contribution is dominated by the two-photon component, the emission state, characterized by the average photon number $\langle n\rangle$ and the second-order correlation function $g^{(2)}(0)$, can be described by the emission probabilities of vacuum, single-photon and two-photon states~\cite{bozzio2022Enhancing, morrison2023singleemitter}
\begin{equation}
    \begin{aligned}
    \label{eq_pho_contribution}
        p_2 & =\frac{g^{(2)}(0)\langle n\rangle^2}{2}, \\
        p_1 & =\langle n\rangle - 2p_2, \\
        p_0 & =1 - p_1 - p_2.
    \end{aligned}
\end{equation}

According to Ref.~\cite{morrison2023singleemitter}, the total gain of the single-photon BB84 protocol in each basis is given by
\begin{equation}
    \label{eq_Qs}
    Q_{x,z} = c_{dt}\sum_{n=0}^\infty p_{n}\left[1-(1-p_{d})(1-\eta_{\rm tot}\eta_{\rm att})^{n}\right],
\end{equation}
where $p_{n}$ is the probability of emitting $n$ photons, $p_{d}$ is the dark count rate of detectors, and $\eta_{\rm tot}=\eta_{\rm cha}\times\eta_{\rm det}$ is the total efficiency of channels and detectors. To reduce the multi-photon events, a pre-attenuation factor $\eta_{\rm att}$ is inserted between the photon source and channel. Taking into account the dead time of detectors, a correction factor $c_{dt}=1/(1+R\tau Q_{x,z})$ is added into the calculation, where $R$ is the repetition rate and $\tau$ is the dead time of detectors. The error gain in each basis can be expressed as
\begin{equation}
\begin{aligned}
    \label{eq_Qes}
    Q_{x,z}^{e} = &c_{dt}\left\{p_{0}p_{d}+\sum_{n=1}^\infty p_{n}p_{\rm mis}\right.\\ &~~~~~~\times\left.\left[1-(1-p_{d})(1-\eta_{\rm tot}\eta_{\rm att})^{n}\right]\right\},
\end{aligned}
\end{equation}
where $p_{\rm mis}$ is the probability of misalignment error rate of the setup.

In this traditional BB84 protocol, Alice encodes $N$ emitted pulses in the $\mathit{Z}$ and $\mathit{X}$ bases with the preparation probabilities $q_{z}$ and ${q_x}=1-q_{z}$ and sends polarization-encoded states to Bob. Bob performs measurements in the rectilinear bases $\{\ket{H},\ket{V}\}$ (as the $\mathit{Z}$ basis) or diagonal bases $\{\ket{+},\ket{-}\}$ (as the $\mathit{X}$ basis), with the measurement probabilities $p_{z}$ and $p_{x}=1-p_{z}$. The data of the $\mathit{Z}$ basis are used for parameter estimation, and those of the $\mathit{X}$ basis are used to extract secure key. Hence, the total number of valid events of the $\mathit{Z}$ and $\mathit{X}$ basis are $n_{z}=Np_{z}q_{z}Q_{z}$ and $n_{x}=Np_{x}q_{x}Q_{x}$. The number of error events of the $\mathit{Z}$ and $\mathit{X}$ basis are $m_{z}=Np_{z}q_{z}Q_{z}^{e}$ and $m_{x}=Np_{x}q_{x}Q_{x}^{e}$. Accordingly, the qubit error rate (QBER) of the $\mathit{Z}$ and $\mathit{X}$ basis are $E_{z}=m_{z}/n_{z}$ and $E_{x}=m_{x}/n_{x}$, respectively.

In this model, we assume that all multi-photon pulses are detected by Bob, offering the expected number of multi-photon events $n_{x(z)}^{\rm mp*}=Np_{x(z)}q_{x(z)}p_{m}$ where $p_{m}\le g^{(2)}(0)\langle n\rangle^2/2$~\cite{waks2002security}. As the number of multi-photon emissions is not observed directly, we use the Chernoff bound to calculate the statistical fluctuations. Given an expected value $x^*$ and failure probability $\varepsilon_{\rm{PE}}$, the upper bound of the observed value can be given by
\begin{equation}
    \label{eq_upperx}
    \bar{x} = x^*+\frac{\beta}{2}+\sqrt{2\beta x^*+\frac{\beta^2}{4}},
\end{equation}
where $\beta=-\ln~\varepsilon_{\rm{PE}}$. Based on \eqref{eq_upperx}, we can calculate the upper bound number of multi-photon events $\overline{n}_{x(z)}^{\rm mp}$ and the lower bound of non-multi-photon events $\underline{n}_{x(z)}^{\rm nmp}=n_{x(z)}-\overline{n}_{x(z)}^{\rm mp}$. We conservatively assume that all errors of the $\mathit{Z}$ basis occur on the received non-multi-photon events, thereby providing the non-multi-photon error rate $E_{z}^{\rm nmp}=m_{z} / \underline{n}_{z}^{\rm nmp}$. We utilize the non-multi-photon events of $\mathit{Z}$ basis to estimate the phase error rate of $\mathit{X}$ basis. The random sampling theorem can be applied to give a upper bound of the phase error rate in the $\mathit{X}$ basis as~\cite{morrison2023singleemitter}
\begin{equation}
    \label{eq_phix_nmp}
    \phi_{x}\le E_{z}^{\rm nmp} + \gamma^{U}(\underline{n}_{x}^{\rm nmp},\underline{n}_{z}^{\rm nmp},E_{z}^{\rm nmp},\varepsilon_{\rm{sec}}/6),
\end{equation}
where  
\begin{equation}
    \begin{aligned}
        \label{eq_gamma1}
        & \gamma^{U}(n,k,\lambda,\epsilon)=\frac{1}{2+2\frac{A^2G}{(n+k)^2}}\times \\
        &~~~~~~\left\{\frac{(1-2\lambda)AG}{n+k} + \sqrt{\frac{A^2G^2}{(n+k)^2}+4\lambda(1-\lambda)G}\right\}, \\
        & A=\max\{n,k\}, \\
        & G=\frac{n+k}{nk}\ln{\frac{n+k}{2\pi nk\lambda(1-\lambda)\epsilon^{2}}}.
    \end{aligned}
\end{equation}

The overall secrecy parameter of the protocol satisfies $\varepsilon_{\rm{sec}}\ge\varepsilon_{\rm{PA}}+\varepsilon_{\rm{PE}}+\varepsilon_{\rm{EC}}$. Here, $\varepsilon_{\rm{PA}}=\varepsilon'$ denotes the failure probability of privacy amplification, $\varepsilon_{\rm{PE}}=2n_{\rm{PE}}\varepsilon'$ denotes the failure probability of parameter estimation where $n_{\rm{PE}} = 2$ is the number of constraints considered in the post-processing procedure, and $\varepsilon_{\rm{EC}}=\varepsilon'$ denotes the failure probability of error correction. Thus, the secrecy coefficient can be set as $\varepsilon_{\rm{sec}}=6\varepsilon'$. With the correction coefficient $\varepsilon_{\rm{cor}}$, the length of secure key can be expressed as~\cite{morrison2023singleemitter, yang2024Highrate}
\begin{equation}
    \label{eq_FSKR_ls}
    \ell=\left\lfloor \underline{n}_{x}^{\rm nmp}[1-h(\phi_{x})]-\lambda_{\rm{EC}}-\log_2\frac{2}{\varepsilon_{\rm{cor}}}-2\log_2\frac{1}{2\varepsilon_{\rm{PA}}} \right\rfloor ,
\end{equation}
where $\lambda_{\rm{EC}}=n_{x}fh(E_{x})$ is the information revealed in the error correction step, $f=1.16$ is the correction efficiency factor, and $h(x)=-x\log_2x-(1-x)\log_2(1-x)$ is the binary Shannon entropy function. The secure key rate (SKR) is then defined as $r=\ell/N$.

\section{Appendix C: Odd cat state as non-classical source for SI-QKD}
In this section, we discuss the feasibility of employing the odd cat state as the non-classical source in our SI-QKD protocol. The cat state is formed by the quantum interference of two coherent states with opposite phases. In the odd-parity form~\cite{wineland2013nobel,li2017cat,he2023fast}, the state can be expressed as

\begin{equation}
    \label{eq_cat}
    |\psi\rangle = \frac{1}{\sqrt{2(1-e^{-2|\alpha|^2})}}(|\alpha\rangle-|-\alpha\rangle),
\end{equation}
where $|\alpha\rangle$ and $|-\alpha\rangle$ represent coherent states with same amplitude and opposite phase. Expanded in the Fock basis, the ideal odd cat state contains only odd photon-number components. 

\begin{figure}[ht]
\centering
\includegraphics[width=\linewidth]{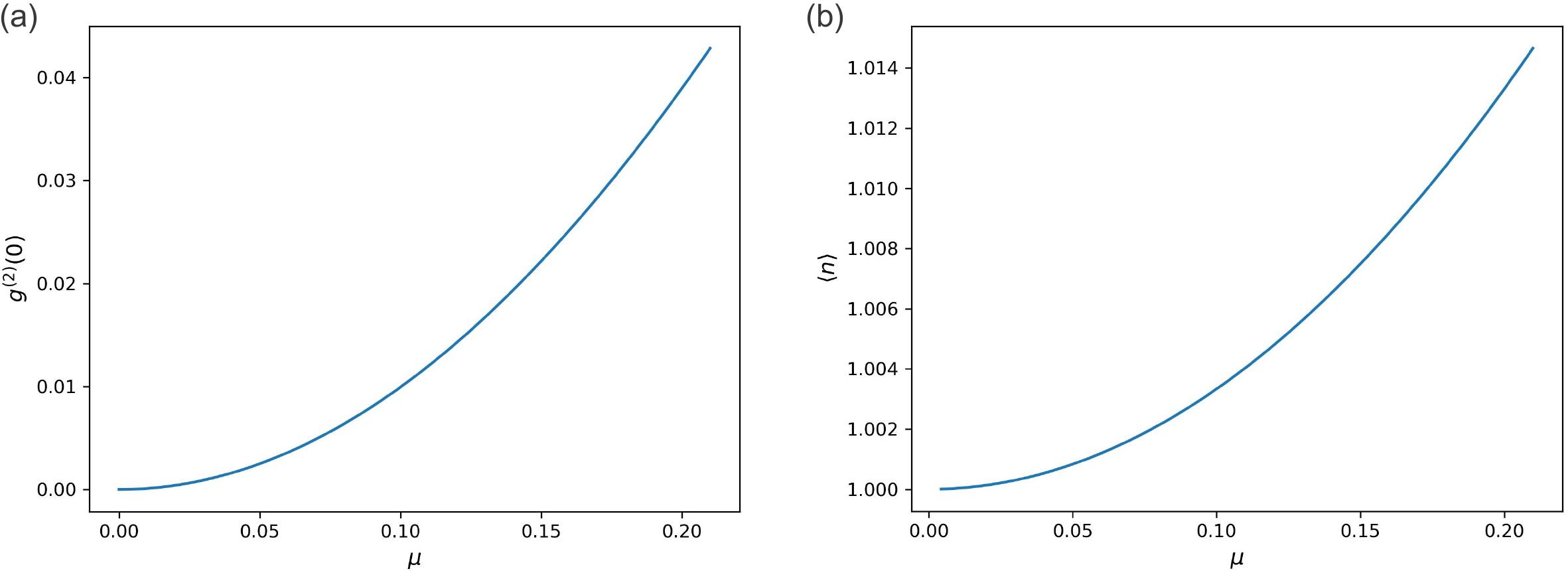}
    \caption{\textbf{Characterization of odd cat state.} 
    (a) Second-order function $g^{(2)}(0)$ as a function of the intensity $\mu=|\alpha|^2$ of the coherent state $|\alpha\rangle$. 
    (b) Average photon number $\langle n\rangle$ of odd cat state as a function of $\mu$. }
    \label{fig-cat-state}
\end{figure}

More generally, we quantify the multi-photon contribution with the second-order correlation function $g^{(2)}(0)$, which can be calculated by

\begin{equation}
    \label{eq_cat_g2}
        g^{(2)}(0) = \frac{\langle \hat{a}^\dagger\hat{a}^\dagger\hat{a}\hat{a}\rangle}{\langle\hat{a}^\dagger\hat{a}\rangle^2} = \left(\frac{1-e^{-2|\alpha|^2}}{1+e^{-2|\alpha|^2}}\right)^2 = \tanh^2(\mu),
\end{equation}
where $\hat{a}$ and $\hat{a}^\dagger$ are the annihilation and creation operators, and $\mu=|\alpha|^2$ is the intensity of the coherent state $\ket{\pm\alpha}$. The average photon number of the odd cat state can be calculated by
\begin{equation}
    \langle n\rangle = \langle\psi|\hat{a}^\dagger\hat{a}|\psi\rangle = \mu \coth(\mu).
\end{equation}
To illustrate the behavior of the odd cat state as a function of $\mu$, we plot the relationship in Fig.~\ref{fig-cat-state}.

For small coherence intensity $\mu$, the Fock-basis expansion of the odd cat state is dominated by the single-photon component, yielding $g^{(2)}(0)\to 0$ and $\langle n\rangle\to 1$. Hence, the odd cat state in the small $\mu$ regime effectively approximate the single-photon state $\ket{1}$ while suppressing multi-photon contributions. This combination of high single-photon purity and pronounced non-classical property makes the odd cat state a promising and practical candidate source for implementing the proposed SI-QKD protocol.

\section{Appendix D: Derivation for the polarization scheme of SI-QKD}

\subsection{D1: Calculation for gains and QBER}
In the setting of SI-QKD, the non-classical source is employed to emit independent optical pulses with high-purity single-photon. We utilize the photon number distribution of single-photon source (SPS) in \eqref{eq_pho_contribution} to characterize each emitted pulse, which can be expressed as 
\begin{equation}
    \label{eq_ini_state}
    \sqrt{p_0}\ket{0}+\sqrt{p_1}\ket{1}+\sqrt{p_2}\ket{2}.
\end{equation}
We assume that the contribution of multi-photon components is dominated by the two-photon term and the imperfection of source is quantified by the second-order correlation function $g^{(2)}(0)$. 

In the polarization scheme, the sequence of pulses are prepared in initial state $\ket{+}$ by a polarizer and a half-wave plate. The optical pulses from two independent SPS are interfered at a polarizing beam slitter (PBS) and then distributed to Alice and Bob. With the initial density matrix $\rho_{c/d}=\left( p_0\ketbra{0}{0} + p_1\ketbra{1}{1} + p_2\ketbra{2}{2} \right)_{c/d}$, the joint density matrix can be explained as
\begin{equation}
    \begin{aligned}
    \label{eq_ini_density}
          &  \rho_{c}\otimes \rho_{d} \\
        =~& p_0^2\ketbra{00}{00}_{cd} + p_0p_1(\ketbra{01}{01}_{cd}+\ketbra{10}{10}_{cd}) \\+& p_0p_2(\ketbra{02}{02}_{cd}+\ketbra{20}{20}_{cd}) 
        + p_1^2\ketbra{11}{11}_{cd} \\+& p_1p_2(\ketbra{12}{12}_{cd}+\ketbra{21}{21}_{cd}) + p_2^2\ketbra{22}{22}_{cd},
    \end{aligned}
\end{equation}
% where $i\in 2\mathbb{Z}+1$ is odd integers and $\ket{12}$ denotes the joint state with 1-photon in time slot $i$ and 2-photon in $i+1$. In the polarization scheme, it corresponds to $\ket{1_+}_{up}\ket{2_+}_{low}$ with one $\ket{+}$-polarized photon in the upper path and two in the lower path. 
where modes $c$ and $d$ are two input ports of PBS. Each term of the state can be considered as an independent pure state. Accordingly, we present the evolution of each component after passing through the PBS in following. The joint state containing 0-photon evolves as
\begin{equation}
    \label{eq_evol_z0}
        |00\rangle_{cd}\xrightarrow{\rm{PBS}}|00\rangle^{HV}_{a}|00\rangle^{HV}_{b}.
\end{equation}
The joint state containing 1-photon evolves as
\begin{equation}
    \begin{aligned}
    \label{eq_evol_z1}
        & |01\rangle_{cd}\xrightarrow{\rm{PBS}} \frac{|01\rangle^{HV}_{a}|00\rangle^{HV}_{b}+|00\rangle^{HV}_{a}|10\rangle^{HV}_{b}}{\sqrt{2}} ,\\
        & |10\rangle_{cd}\xrightarrow{\rm{PBS}} \frac{|10\rangle^{HV}_{a}|00\rangle^{HV}_{b}+|00\rangle^{HV}_{a}|01\rangle^{HV}_{b}}{\sqrt{2}} .\\
    \end{aligned}
\end{equation}
The joint state containing 2-photon evolves as
\begin{equation*}
    \begin{aligned}
    \label{eq_evol_z2}
        & |02\rangle_{cd}\xrightarrow{\rm{PBS}} \\&\frac{|02\rangle^{HV}_{a}|00\rangle^{HV}_{b}+\sqrt{2}|01\rangle^{HV}_{a}|10\rangle^{HV}_{b}+|00\rangle^{HV}_{a}|20\rangle^{HV}_{b}}{2} ,\\
        & |20\rangle_{cd}\xrightarrow{\rm{PBS}} \\&\frac{|20\rangle^{HV}_{a}|00\rangle^{HV}_{b}+\sqrt{2}|10\rangle^{HV}_{a}|01\rangle^{HV}_{b}+|00\rangle^{HV}_{a}|02\rangle^{HV}_{b}}{2} ,\\
        & |11\rangle_{cd}\xrightarrow{\rm{PBS}} \frac{1}{2}\Big(|11\rangle^{HV}_{a}|00\rangle^{HV}_{b} + |10\rangle^{HV}_{a}|10\rangle^{HV}_{b} \\&~~~~~~~~~~~~~~~~+ |01\rangle^{HV}_{a}|01\rangle^{HV}_{b} + |00\rangle^{HV}_{a}|11\rangle^{HV}_{b}\Big) .\\
    \end{aligned}
\end{equation*}
The joint state containing 3-photon evolves as
\begin{equation}
    \begin{aligned}
    \label{eq_evol_z3}
        & |12\rangle_{cd}\xrightarrow{\rm{PBS}} \\&\frac{1}{2\sqrt{2}} \Big(|12\rangle^{HV}_{a}|00\rangle^{HV}_{b} + \sqrt{2}|11\rangle^{HV}_{a}|10\rangle^{HV}_{b} + |10\rangle^{HV}_{a}|20\rangle^{HV}_{b} \\&\hspace{1em}+|02\rangle^{HV}_{a}|01\rangle^{HV}_{b} + \sqrt{2}|01\rangle^{HV}_{a}|11\rangle^{HV}_{b} + |00\rangle^{HV}_{a}|21\rangle^{HV}_{b}\Big) ,\\
        & |21\rangle_{cd}\xrightarrow{\rm{PBS}} \\&\frac{1}{2\sqrt{2}} \Big(|21\rangle^{HV}_{a}|00\rangle^{HV}_{b} + \sqrt{2}|11\rangle^{HV}_{a}|01\rangle^{HV}_{b} + |01\rangle^{HV}_{a}|02\rangle^{HV}_{b} \\&\hspace{1em}+ |20\rangle^{HV}_{a}|10\rangle^{HV}_{b} + \sqrt{2}|10\rangle^{HV}_{a}|11\rangle^{HV}_{b}+|00\rangle^{HV}_{a}|12\rangle^{HV}_{b}\Big) .\\
    \end{aligned}
\end{equation}
The joint state containing 4-photon evolves as
\begin{equation}
    \begin{aligned}
    \label{eq_evol_z4}
        & |22\rangle_{cd}\xrightarrow{\rm{PBS}} \\&\frac{1}{4} \left(|22\rangle^{HV}_{a}|00\rangle^{HV}_{b} + \sqrt{2}|21\rangle^{HV}_{a}|10\rangle^{HV}_{b} + |20\rangle^{HV}_{a}|20\rangle^{HV}_{b} \right.\\
        & \hspace{1em} \left. + \sqrt{2}|12\rangle^{HV}_{a}|01\rangle^{HV}_{b} + 2|11\rangle^{HV}_{a}|11\rangle^{HV}_{b} + \sqrt{2}|10\rangle^{HV}_{a}|21\rangle^{HV}_{b}  \right.\\
        & \hspace{1em} \left. + |02\rangle^{HV}_{a}|02\rangle^{HV}_{b}+\sqrt{2}|01\rangle^{HV}_{a}|12\rangle^{HV}_{b}+|00\rangle^{HV}_{a}|22\rangle^{HV}_{b}\right) .
    \end{aligned}
\end{equation}
The measurement operators of click and no-click can be written as
\begin{equation}
    \begin{aligned}
        \label{eq_Fc_Fnc}
        \hat{F}_c & = \hat{\mathrm{I}}-\sum_{n=0}^\infty (1-p_d)(1-\eta)^n|n\rangle\langle n| \\&= \sum_{n=0}^\infty \left[ 1-(1-p_d)(1-\eta)^n \right]|n\rangle\langle n|,\\
        \hat{F}_{nc} & = \hat{\mathrm{I}}-\hat{F}_c = \sum_{n=0}^\infty(1-p_d)(1-\eta)^n|n\rangle\langle n| .
    \end{aligned}
\end{equation}
where $p_{d}$ is the dark count rate, $\eta$ is the total efficiency for the measurement basis, $\hat{\mathrm{I}}=\sum_{n=0}^{\infty}\ketbra{n}{n}$ is the identical operator of photon number space, and $\ket{n}$ is the Fock state with $n$-photon.

In the polarization scheme, a valid detection event is defined as Alice and Bob obtain clicks from the same measurement basis in the same time slot. If more than one detector obtain clicks, they randomly chose one as the effective click. For the rectilinear basis, detections from $\ket{H}$ and $\ket{V}$ are assigned as bit 0 and 1 of the $\mathit{Z}$ basis, respectively. For the diagonal basis, detections from $\ket{+}$ and $\ket{-}$ are assigned as bit 0 and 1 of the $\mathit{X}$ basis, respectively. Thereby, a correct detection event is defined as Alice and Bob obtain clicks from the same type of detector. 

For each photon-number component of the quantum state in Eq.~(\ref{eq_ini_density}), we calculate its contribution to the valid detection events in the following formulas. For the component containing 0-photon, the correct and error gains are given by
\begin{equation}
    \begin{aligned}
    \label{eq_Q0}
        Q_{0}^{c}=Q_{0}^{e}=2p_{0}^2p_{d}^2\left(1-\frac{1}{2}p_{d}\right)^2. 
    \end{aligned}
\end{equation}
For the component containing 1-photon, the correct and error gains are given by 
\begin{equation}
    \begin{aligned}
    \label{eq_Q1}
        & Q_{1}^{c}=Q_{1}^{e}=2p_{0}p_{1}p_{d}\left(1-\frac{1}{2}p_{d}\right)\left[1-(1-p_{d})^2(1-\eta)\right]. \\
    \end{aligned}
\end{equation}
For the component containing 2-photon, the correct and error gains are given by 
\begin{equation}
    \begin{aligned}
    \label{eq_Q2}
        & Q_{2}^{c}=2p_{0}p_{2}p_{d}\left(1-\frac{1}{2}p_{d}\right)\left[1-(1-p_{d})^2(1-\eta)^2\right] \\&\hspace{2em}+ \frac{p_{1}^2}{2}\left[1-(1-p_{d})^2(1-\eta)\right]^2 ,\\
        & Q_{2}^{e}=p_{0}p_{2}\left[1-(1-p_{d})^2(1-\eta)\right]^2 \\&\hspace{2em}+ p_{1}^2p_{d}\left(1-\frac{1}{2}p_{d}\right)\left[1-(1-p_{d})^2(1-\eta)^2\right] .\\
    \end{aligned}
\end{equation}
For the component containing 3-photon, the correct and error gains are given by 
\begin{equation}
    \begin{aligned}
    \label{eq_Q3}
        & Q_{3}^{c}=p_{1}p_{2}\left[1-(1-p_{d})^2(1-\eta)\right]\left[1-(1-p_{d})^2(1-\eta)^2\right] ,\\
        & Q_{3}^{e}=\frac{p_{1}p_{2}}{2}\Big[1+2p_{d}-p_{d}^2-(1-p_{d})^2(1-\eta)(2-\eta) \\&\hspace{2em}+ \left(1-4p_{d}+2p_{d}^2\right)(1-p_{d})^2(1-\eta)^3\Big] .\\
    \end{aligned}
\end{equation}
For the component containing 4-photon, the correct and error gains are given by 
\begin{equation}
    \begin{aligned}
    \label{eq_Q4}
        & Q_{4}^{c}=\frac{p_{2}^2}{2}\left[1-(1-p_{d})^2(1-\eta)^2\right]^2 ,\\
        & Q_{4}^{e}=\frac{p_{2}^2}{8}\Big[3+2p_{d}-p_{d}^2-2(1-p_{d})^2(1-\eta)\left(3-3\eta+2\eta^2\right) \\&\hspace{2em}+ (3-8p_{d}+4p_{d}^2)(1-p_{d})^2(1-\eta)^4\Big] . 
    \end{aligned}
\end{equation}
Substitute the efficiency of the $\mathit{Z}$ and $\mathit{X}$ basis, where $\eta_{z}=p_{z}\times\eta_{\rm det}\times\eta_{\rm cha}$ and $\eta_{x}=p_{x}\times\eta_{\rm det}\times\eta_{\rm cha}$, into formulas in Eqs.~(\ref{eq_Q0})-(\ref{eq_Q4}) to obtain the gains and QBER. Consequently, the correct gain, error gain, total gain and QBER can be expressed as
\begin{equation}
    \begin{aligned}
    \label{eq_QEz}
        & Q_{z(x)}^{c} = \sum_{i=1}^{4}Q_{i}^{c}(
        \eta_{z(x)}),~~Q_{z(x)}^{e} = \sum_{i=1}^{4}Q_{i}^{e}(\eta_{z(x)}), \\
        & Q_{z(x)} = Q_{z(x)}^{c} + Q_{z(x)}^{e},\\&       Q_{z(x)}E_{z(x)} = e_{d}Q_{z(x)}^{c}+(1-e_{d})Q_{z(x)}^{e}.
    \end{aligned}
\end{equation}
Assuming ideal detectors with $p_d=0$ and $e_d=0$, we conclude
\begin{equation}
    \begin{aligned}
    \label{eq_QzEz_ideal}
        & Q_{z(x)}^{c}=\frac{\langle n \rangle^2\eta_{z(x)}^2}{2}~, ~Q_{z(x)}^{e}=\frac{g^{(2)}(0)\langle n \rangle^2\eta_{z(x)}^2}{2}~,\\
        & Q_{z(x)}=\left(1+g^{(2)}(0)\right)\frac{\langle n\rangle^2\eta_{z(x)}^2}{2}~, \\& E_{z}=E_{x}=\frac{g^{(2)}(0)}{1+g^{(2)}(0)}~.
    \end{aligned}
\end{equation}

\subsection{D2: Calculation for secure ket rate}
According to the entropy uncertainty relation, the secret key length of the polarization scheme in the finite-size regime can be written as~\cite{tomamichel2012Tight, yin2019Finitekey}
\begin{equation}
    \label{eq_FSKR_lz}
    \ell = n_{z}[1-h(\phi_{z})]-\lambda_{\rm{EC}}-\log_2\frac{2}{\varepsilon_{\rm{cor}}}-2\log_2\frac{1}{\varepsilon_{\rm{sec}}},
\end{equation}
where $n_{z}=Np_{z}^2Q_{z}$ is the number of valid events in the $\mathit{Z}$ basis, $\phi_{z}$ is the phase error rate of the $\mathit{Z}$ basis, and $\lambda_{\rm{EC}}=n_{z}fh(E_{z})$ is the information leaked during error correction. $\varepsilon_{\rm{sec}}$ and $\varepsilon_{\rm{cor}}$ are the security coefficients regarding the secrecy and correctness, respectively, and $h(x)=-x\log_2x-(1-x)\log_2(1-x)$ is the binary Shannon entropy function. Then the finite SKR is defined as $r=\ell/N$. 

Here, we extract random data from the $\mathit{Z}$ basis to generate the secure key and the remaining for error correction. All data in the $\mathit{X}$ basis are revealed for parameter estimation. Based on the random sampling theorem without replacement problem, the phase error rate of the $\mathit{Z}$ basis can be bounded by~\cite{yin2020tight}
\begin{equation}
    \label{eq_phiz}
    \phi_{z}\le E_{x} + \gamma^{U}(n_{z},n_{x},E_{x},\epsilon),
\end{equation}
where the statistical method based on Chernoff bound is the same as \eqref{eq_gamma1}. By composing the failure probability due to parameter estimation, the protocol with security coefficient $\varepsilon_{\rm{sec}}$ takes $\epsilon=\varepsilon_{\rm{sec}}/2$.

\section{Appendix E: Experimental feasibility}

The practical viability of our SI-QKD protocol is primarily determined by the indistinguishability of independently generated single photons and the long-distance stability of quantum channels. In particular, high-visibility interference between independent optical pulses requires simultaneous control of spectral overlap, temporal synchronization, and polarization alignment. 

A central challenge lies in reconciling photon indistinguishability with source brightness. In practical SPSs, spectral diffusion and pure dephasing often compromise the temporal coherence of successive photons. This limitation can be effectively mitigated by employing narrow-linewidth spectral filtering, which suppresses spectral broadening and restores near-transform-limited emission. Although such filtering inevitably reduces the photon emission rate, recent experimental progress in Ref.~\cite{you2022quantum} shows that high interference visibility can still be achieved simultaneously with sufficient brightness for long-distance quantum interference, even after transmission over 300 km of optical fiber. Importantly, their analysis confirms that residual multi-photon contributions and spectral mismatch do not fundamentally limit interference visibility in QKD. 

In our polarization SI-QKD protocol, the requirements on photon indistinguishability are further relaxed compared to phase-encoded or time-phase encoded schemes. Since interference occurs at a PBS and relies on polarization overlap, the protocol is less sensitive to frequency detunings and phase noise. Consequently, the trade-off between brightness and indistinguishability has a reduced impact on the achievable SKR. 

Recent breakthroughs in quantum light source also demonstrates that high-purity single-photon emitters can simultaneously provide high brightness and  near-unity indistinguishability. Specially, engineered quantum-dot-cavity systems have achieved both high indistinguishability and system efficiency~\cite{ding2025high}, while versatile computing platforms enable high-visibility interference across reconfigurable networks with preserved coherence~\cite{maring2024versatile}. Furthermore, multi-photon interference and entanglement generation have been realized in both integrated photonic platforms~\cite{chen2024heralded} and complex linear-optical networks~\cite{cao2024photonic}. These advances provide the necessary indistinguishability and brightness required in our protocol. 

Based on these technological solutions, precise temporal synchronization and polarization stability are attainable, specifically sub-nanosecond clock alignment with tunable optical delay lines and real-time dynamic polarization control systems that compensate for fiber-induced rotations. Collectively, these results offer a solid experimental foundation to the high-rate quantum communication over long-distance transmission in polarization-encoded implementations.

\begin{figure}[ht]
\centering
\includegraphics[width=\linewidth]{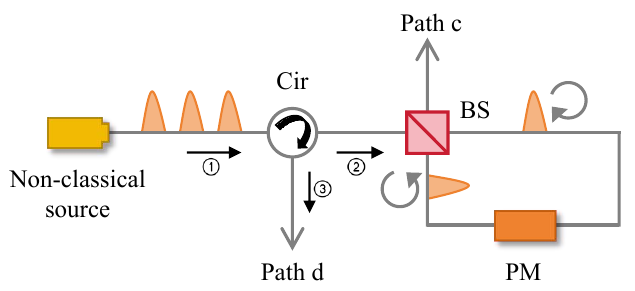}
    \caption{\textbf{The experimental setup of optical switch.} Single photons enter the Sagnac interferometer via the circulator (Cir) and beam splitter (BS). An electro-optic phase modulator (PM) applies periodic phase shifts at specific time slots. Depending on the relative phase ($\pi$ or 0), photons from odd and even time slots are routed to separate output ports ($c$ and $d$). }
    \label{fig-optical switch}
\end{figure}

Notably, the protocol can be realized with a single non-classical source integrated with an active optical switch. State-of-art technology has enabled high-speed optical switching utilizing the Sagnac interferometers and electro-optic phase modulators (PM). This methodology can achieve GHz-level repetition rates, which is widely adopted in high-rate QKD systems~\cite{tang2023time,wang2026time}. As illustrated in Fig.~\ref{fig-optical switch}, single-photon pulses are guided into the Sagnac interferometer through a beam splitter (BS), where each pulse is split into clockwise ($\circlearrowright$) and counterclockwise ($\circlearrowleft$) components. Due to the asymmetric propagation, the two components arrive at the PM with a relative time delay $\Delta$, which is set to half of the single-photon repetition period. Assuming that the single photon along the counterclockwise ($\circlearrowleft$) path in an odd pulse arrives at time $t_0$, we apply a $\pi$ phase shift at time slots $t=t_0+(1+4i)\Delta$ (i=0,1,2,...). Under this modulation, only the clockwise ($\circlearrowright$) component of the odd pulse acquires a $\pi$ phase, while both components of the even pulse remain unmodulated. As a result, a relative phase difference of $\pi$ (0) is introduced for odd (even) pulses between the two paths of Sagnac loop. Through the interference at the BS, this scheme allows the odd and even pulses to deterministically routed to paths $c$ and $d$, respectively. These outputs are subsequently directed to the input ports of the PBS for further interference.

While the scheme can be further simplified by replacing the active optical switch with a passive 50:50 BS, this architecture introduces a significant compromise in source utilization efficiency. In this configuration, each input single-photon pulse is probabilistically routed, which can be expressed as
\begin{equation}
    |+\rangle_{in}\xrightarrow{\rm BS}\frac{1}{\sqrt{2}}\left(|+\rangle_{c}+|+\rangle_{d}\right).
\end{equation}
By introducing a time delay in path $c$, adjacent pulses are temporally aligned such that the $i$-th time slot of path $c$ coincides with the $i+1$-th time slot of path $d$. Thus, the interference evolution at the PBS can be described as
\begin{equation}
\begin{aligned}
    \frac{1}{2}(|+\rangle_{c}\otimes|+\rangle_{d})\xrightarrow{\rm PBS}&\frac{1}{4}\Big(|HH\rangle_{ab}+|VV\rangle_{ab}\\&~+|HV\rangle_{aa}+|VH\rangle_{bb}\Big),
\end{aligned}
\end{equation}
where modes $a$ and $b$ denote the output ports guiding to Alice and Bob, respectively. While this passive approach eliminates the need for high-speed modulation and reduces experimental complexity, it results in a 50\% reduction in effective interference events and decrease the system efficiency.

\section{Appendix F: Simulation results}

\subsection{F1: System parameters}
\begin{table}[b]
    \centering
    \caption{Simulation parameters. }
    \begin{tabular}{c c c}
    \hline
    \hline
    \textbf{Description} & \textbf{Parameter} & \textbf{Value} \\ 
    \hline
    Detection efficiency            & $\eta_{\rm{det}}$        & 0.8              \\
    Average fiber loss              & $\alpha$                 & 0.16 \rm{dB/km}  \\
    Error correction efficiency   & f                        & 1.16             \\
    Correctness failure probability & $\varepsilon_{\rm{cor}}$ & $10^{-15}$        \\
    Secrecy failure probability     & $\varepsilon_{\rm{sec}}$ & $10^{-10}$        \\
    \hline
    \hline
    \end{tabular}
    \label{tab:simulation parameter}
\end{table}

\begin{figure*}[ht]
\centering
\includegraphics[width=\linewidth]{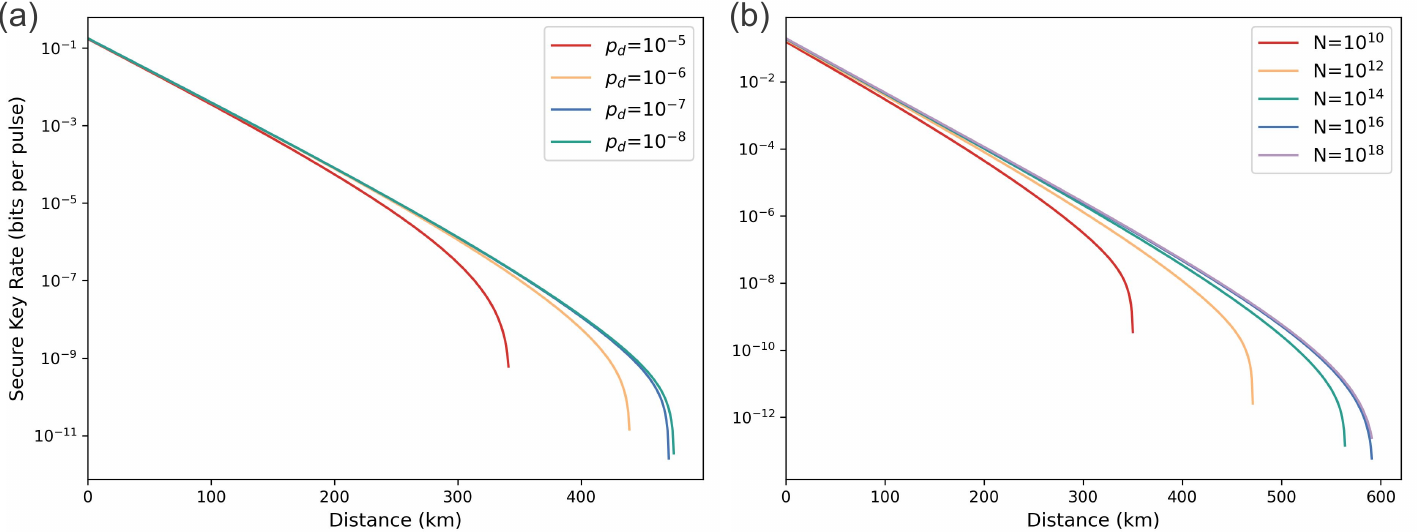}
    \caption{\textbf{The SKR under different condition.} Finite secure key rate versus transmission distance with fixed misalignment error $e_d=0.01$ and second-order coherence $g^{(2)}(0)=0.01$.
    (a) The dark count rate $p_d$ ranging from $10^{-5}$ to $10^{-8}$ with $N=10^{12}$.
    (b) The date size $N$ ranging from $10^{10}$ to $10^{18}$ with $p_d=10^{-7}$. }
    \label{fig-performance}
\end{figure*}

Here we utilize the parameters listed in Table~\ref{tab:simulation parameter} to simulate the finite SKR of our protocol. The channel efficiency from Charlie to Alice (Bob) is $\eta_{a}=10^{-\alpha l_{a}/10}$ ($\eta_{b}=10^{-\alpha l_{b}/10}$), where $l_{a}$ ($l_{b}$) is the distance between the source and Alice (Bob). For simplicity, we assume symmetric channels for Alice and Bob, hence the total channel length is $l=l_{a}+l_{b}=2l_{a}$ and the channel transmission efficiency is $\eta_{a} = \eta_{b} = \eta_{\rm{cha}}=10^{-\alpha l/20}$. Nevertheless, our SI-QKD protocol remains applicable to asymmetric channels.

\subsection{F2: Performance analysis}

To evaluate the performance of the proposed polarization scheme, we investigate the dependence of SKR on the detector dark count rate $p_d$, as shown in Fig.~\ref{fig-performance}(a). With fixed misalignment error $e_d=0.01$, data size $N=10^{12}$ and the second-order coherence $g^{(2)}(0)=0.01$, the mean photon number $\langle n\rangle$ and the measurement probability of $\mathit{Z}$ basis $p_z$ are optimized for each distance to maximize the SKR. The simulation results exhibit an achievable transmission distance exceeding 300 km under a high dark count rate of $p_d=10^{-5}$. As $p_d$ is suppressed to $10^{-8}$, the SKR eventually saturates, where the wrong detection events mainly account to the intrinsic alignment errors and the multi-photon components. 

Fig.~\ref{fig-performance}(b) further illustrates the SKR under varying numbers of emitted pulses $N$, considering the dark count rate $p_d=10^{-7}$, the misalignment error $e_d=0.01$ and the source quality $g^{(2)}(0)=0.01$. For $N\ge10^{16}$, the key rate stabilizes and converges towards the asymptotic limit. This demonstrates that the proposed protocol maintains high performance under realistic data accumulation times, confirming its feasibility for practical experimental implementations.

% Bibliography
% \bibliography{supplement}

%\bibliographystyle{modified-apsrev4-2_new}
%\bibliography{sample.bib}% Produces the bibliography via BibTeX.

%bst derived from apsrev4-2.bst (revtex)
%Control: key (0)
%Control: author (72) initials jnrlst
%Control: editor formatted (1) identically to author
%Control: production of article title (1) required
%Control: page (0) single
%Control: year (1) truncated
%Control: production of eprint (0) enabled
%

\end{document}